\documentclass[aps,twocolumn,superscriptaddress,showpacs,preprintnumbers,amsmath,amssymb]{revtex4}

\usepackage{graphicx}
\usepackage{textcomp}   
\usepackage[scaled]{helvet}  
\usepackage{amsmath,amssymb}  
\usepackage{type1cm}
\usepackage{bm}  
\usepackage{color}
\usepackage{ulem}
\usepackage{comment}

\begin{document}

\title{Casimir Force for the ${\mathbb C}P^{N-1}$  Model}

\author{Antonino Flachi} 
\affiliation{Department of Physics \& Research and Education Center for Natural Sciences, Keio University, 4-1-1 Hiyoshi, Kanagawa 223-8521, Japan}

\author{Muneto Nitta} 
\affiliation{Department of Physics \& Research and Education Center for Natural Sciences, Keio University, 4-1-1 Hiyoshi, Kanagawa 223-8521, Japan}

\author{Satoshi Takada} 
\affiliation{Institute of Engineering, Tokyo University of Agriculture and Technology, 2-24-16, Naka-cho, Koganei, Tokyo 184-8588, Japan}
\affiliation{Earthquake Research Institute, The University of Tokyo, 1-1-1 Yayoi, Bunkyo-ku, Tokyo 113-0032, Japan}

\author{Ryosuke Yoshii} 
\affiliation{Department of Physics, Chuo University, 1-13-27 Kasuga, Bunkyo-ku, Tokyo 112-8551, Japan}
\affiliation{Research and Education Center for Natural Sciences, Keio University, 4-1-1 Hiyoshi, Kanagawa 223-8521, Japan}

\date{\today}
\begin{abstract}
In this work, we derive exact self-consistent solutions 
to the gap equations of
the $\mathbb{C}P^{N-1}$ model on a finite interval with Dirichlet boundary conditions in the large-$N$ approximation. The solution reproduce the confining phase in the infinite system by taking the appropriate limit. 
We compute the vacuum energy and the Casimir force and observe that the sign of the force is always attractive. 
\end{abstract}

\maketitle

\section{Introduction}
The Casimir force is a peculiar effect originating from 
the deformation of the quantum vacuum of the electromagnetic field caused by the presence of boundaries \cite{casimir}.
This force decays as inverse powers of the size of the system, but dominates for small distances, making the Casimir effect of paramount importance for nanotechnology applications. The original finding of Casimir showed an attractive force for the electromagnetic field between two parallel, perfectly conducting plates and this
was experimentally detected (unambiguously for the first time) about fifty years after Casimir's original finding \cite{Experiments}. 

Another interesting direction, recently resurrected in \cite{simpson}, is the possibility that, by appropriately modelling the interior of charged particles, the electromagnetic repulsion could be stabilised by an attractive quantum vacuum force. This idea of modelling the electron by using quantum vacuum effects to balance Coulomb repulsion, originally due to Casimir \cite{mouse},  had to be rejected once explicit calculations for spheres \cite{boyer} revealed a repulsive force and showed that at least in the simplest version, this Casimir stabilisation mechanism did not work. The recent analyses of \cite{simpson} seem to have found a way around it.

A number of works (see for example refs.~\cite{Munday,KennethBachas,Schaden,Asorey}) analyzed the question on how to control the sign of the force and it became clear that imposing \textit{ad-hoc} boundary conditions may lead to a change in the sign of the force, turning the issue of the sign into an issue on how to dynamically induce changes in the boundary conditions. A partial answer to this question has been recently proposed in \cite{Flachi:2017cdo}, where it was shown that using an interacting fermion field theory of the Nambu-Jona Lasinio (or chiral Gross-Neveu) class, allows to realize a sign-flip in the force. The idea behind \cite{Flachi:2017cdo} is that the standard attractive Casimir contribution that dominates the vacuum energy for negligible coupling is opposed by a contribution to the effective action from the condensate. The latter produces a repulsive force that competes with the former when the coupling grows larger. It is due to this competition that a flip in the force is generated. Physically, the boundary conditions are altered dynamically by the condensate being localized close to the boundaries that leads to an effective deformation of the boundary conditions and leads to a change in the force.

One important question left for clarification concerns the universality of this mechanism. In other words, whether the same sign-change in the force can be achieved in other systems featuring symmetry breaking, e.g. interacting bosonic systems. This is the question we wish to address in this paper and with this in mind we consider the scalar cousin of the interacting fermion model of Ref.~\cite{Flachi:2017cdo}, that is the ${\mathbb C}P^{N-1}$ model. 
Previous relevant work is that of Ref.~\cite{Schmidt} that looks at an interacting $\lambda \phi^4$ scalar theory. 
The non-perturbative effect on the Casimir force due to the presence of the dynamical topological defects was also discussed in Ref.\ \cite{Chernodub}. 
More recently, the Casimir force for the Yang- Mills theory was also investigated \cite{Chernodub2}. 

The ${\mathbb C}P^{N-1}$ model 
\cite{Eichenherr:1978qa, Golo:1978dd, Cremmer:1978bh} 
in 1+1 dimensions 
has 
a long history due to the
similarities between the sigma model 
and Yang-Mills theory in 3+1 dimensions: 
dynamical mass gap, asymptotic freedom and instantons 
\cite{Polyakov:1975rr, Polyakov:1975yp, Bardeen:1976zh, Brezin:1976qa, DAdda:1978vbw, DAdda:1978etr, Witten:1978bc,Novikov:1984ac}. 
The 1+1 dimensional ${\mathbb C}P^{N-1}$ model also appears as 
 a world-sheet theory of a non-Abelian vortex string 
 in a 3+1 dimensional $U(N)$ gauge theory 
with $N$ scalar fields in the fundamental representation 
\cite{Hanany:2003hp, Auzzi:2003fs, Eto:2005yh}  
(see Refs.~\cite{Tong:2005un,Eto:2006pg,Shifman:2007ce,Tong:2008qd}
for review),  
yielding a nontrivial relation between 
the ${\mathbb C}P^{N-1}$ model and the $U(N)$ gauge theory 
\cite{Hanany:2004ea,Shifman:2004dr}. 
The $\mathbb{C}P^{N-1}$ model on a finite space 
was studied before on a finite interval 
\cite{Milekhin:2012ca,Bolognesi:2016zjp,
Milekhin:2016fai,Pavshinkin:2017kwz} 
as well as on a ring \cite{Monin:2015xwa,Monin:2016vah}, 
describing a closed string as well as 
an open string ending on some boundary, respectively. 
A finite system in 2+1 dimensions, a disc system with the Dirichlet boundary condition at the edge (a circle), was also investigated \cite{Pikalov}. 
In Ref.~\cite{Milekhin:2012ca} 
a phase transition between 
confining (unbroken) phase 
for a larger system
and Higgs (broken) phase
for a smaller system 
was found, 
although a constant configuration was assumed 
inconsistently with the presence of boundaries.
Spatially varying configurations are consistent with the presence of 
boundaries and these were derived numerically \cite{Bolognesi:2016zjp}.
Constant configurations were later  justified by 
changing the boundary conditions  
\cite{Milekhin:2016fai,Pavshinkin:2017kwz}. 
A similar phase transition was also found for the case of a ring 
\cite{Monin:2015xwa,Monin:2016vah}.

Here, we shall work with the ${\mathbb C}P^{N-1}$ model 
on a finite interval and compute the vacuum energy for such a system with Dirichlet boundary conditions imposed at the edges of the interval. 
Analogously to the earlier findings of Ref.~\cite{Flachi:2017cdo,Schmidt}, the force in this case too shows a change in sign generated by the same competition between the (large-$N$) contribution of the Casimir energy and that of the condensate.
The calculations are carried out by adopting a recently developed mapping between the 
$\mathbb{C}P^{N-1}$ model 
and the Gross-Neveu model \cite{Nitta:2017uog}, where 
self-consistent exact solutions to the gap equations of 
the $\mathbb{C}P^{N-1}$ on the infinite line or with periodic boundary conditions were obtained.
Here, we derive, for the first time, exact self-consistent solutions 
to the gap equations for
the $\mathbb{C}P^{N-1}$ model on a finite interval 
with Diriclet boundary conditions
in the large-$N$ limit.
We find that the present solution reproduces the confining solution by taking the appropriate limit. 
We calculate the quantum vacuum force for the obtained solution and find that the force is always attractive 
and we also show that the L\"usher term enhances for the larger interactions.

\section{Model, method, and solutions}
The ${\mathbb C}P^{N-1}$ model is defined by the following action
\begin{equation}
S=\int dtdx \left[(D_\mu n_i)^\ast (D^\mu n_i)-\lambda (n_i^\ast n_i-r)\right], 
\label{action}
\end{equation}
with the $n^i$ ($i=1,\cdots, N$) are complex scalar fields, $D_\mu=\partial_\mu-iA_\mu$, and $\lambda(x)$ is a Lagrange multiplier; $\lambda$ is the mass gap function and $\sigma$ represents the Higgs field. 

Decomposing $n^i$ into $n^0=\sigma$, $n^i=\tau^i$ ($i=2,\cdots, N$) and integrating out the $\tau$ fields we arrive at the following field equations
\begin{align}
&[-\partial_x^2+\lambda(x)]f_n(x)=\omega^2_n f_n(x),\label{eq1}\\
&\frac{N}{2}\sum_n\frac{f_n^2}{\omega_n}+\sigma(x)^2-r=0,\label{eq2}\\
&\partial_x^2\sigma(x)-\lambda(x)\sigma(x)=0, \label{eq3}
\end{align}
with the functions $f_n$ defining the modes and ${\omega_n}$ the eigenfrequencies.
Here we assume the model to live on the interval $[-L/2, L/2]$ with Dirichlet boundary condition at the edges, similarly to Ref.~\cite{Bolognesi:2016zjp}.

The solutions can be found exactly, by using a mapping between the Gross-Neveu model and the ${\mathbb C}P^{N-1}$ model as discussed in Ref.~\cite{Nitta:2017uog}. The map is implemented by the following prescription: 
\begin{eqnarray}
&\lambda=\Delta^2+\partial_x \Delta, \label{lambdamap}\\
&\sigma=A\exp[\int^x_0\Delta(y)dy], \label{sigmamap}
\end{eqnarray}
where $A$ is an integration constant, $\Delta$ is a gap function obeying the Bogoliubov-de Gennes equation
\begin{align}
&\left(
\begin{array}{cc}
0 & \partial_x+\Delta\\
-\partial_x+\Delta & 0
\end{array}
\right)
\left(
\begin{array}{c}
f_n\\
g_n
\end{array}
\right)
=\omega_n
\left(
\begin{array}{c}
f_n\\
g_n
\end{array}
\right), 
\label{BdG}
\end{align}
and the gap equation
\begin{align}
\Delta=\frac{N}{2r}\sum_{\omega_n\ge 0} f_n g_n.
\label{gap}
\end{align}
These equations follow from the Gross-Neveu Lagrangian in the mean field approximation. In other words, one can re-write the equations for the ${\mathbb C}P^{N-1}$ model as those in the Gross-Neveu model by using the above mapping. At this point, it is possible to take advantage of the known exact solution for the Gross-Neveu model to find exact solutions for the ${\mathbb C}P^{N-1}$ model.

Using the same mapping, one can express the total energy in the ${\mathbb C}P^{N-1}$ model in terms of the auxiliary field $\Delta$,
\begin{align}
E&=N\sum_n\omega_n-r\int dx \lambda (x)\nonumber\\
&=N\sum_n\omega_n-r\int dx \left(\Delta^2+\partial_x \Delta\right). 
\label{energy}
\end{align}
Here the surface term $\left.\sigma\partial_x\sigma\right|_{-L/2}^{L/2}$ is dropped in the left hand side, 
which is, for example, justified in the cutoff renormalization scheme: 
one can rewrite the term by $\sigma\partial_x \sigma=N\sum f_n\partial_xf_n/\omega_n$ [Eq.\ (\ref{eq2})] 
and both $f_n$ and $\partial_x f_n$ regularly vanish towards the boundary and thus $\sigma\partial_x \sigma$ also vanishes at the boundary as long as we have the energy cutoff (derivative and summation is commutable). 
Though the cutoff renormalization scheme is used for this argument, the result does not depend on the renormalization scheme.

In the case of the Gross-Neveu model, we have found two classes of self-consistent solutions consistent with the boundary conditions (see Ref.~\cite{Flachi:2017cdo} for details): one class corresponding to BCS-type solutions, 
\begin{equation}
\Delta_{\mathrm{BCS}}=\frac{2\kappa}{\mathrm{sn}(2\kappa x,\nu)},
\label{deltaBCS}
\end{equation} 
and the other corresponding to  normal-type solutions, 
\begin{equation}
\Delta_{\mathrm{Normal}}=-\kappa \frac{\mathrm{cn}(\kappa x+\mathbf{K},\nu)}{\mathrm{sn}(\kappa x+\mathbf{K},\nu)}.
\label{deltanormal}
\end{equation} 
Above we have indicated with $\mathrm{sn}$, $\mathrm{cn}$, and $\mathrm{dn}$ the Jacobi's elliptic functions. The quantity $\nu$ is a constant and defines the elliptic parameter.
Also, we have defined $2\mathbf{K}(\nu)/L\equiv \kappa$, where $\mathbf{K}(\nu)$ is a complete elliptic integral of the first kind. Notice that in the limit $L\rightarrow \infty$ and $\nu\rightarrow 1$ the two solutions reproduce respectively the BCS solution $\Delta=\text{const}$ and the normal solution $\Delta=0$.

\begin{figure}
\includegraphics[width=20pc]{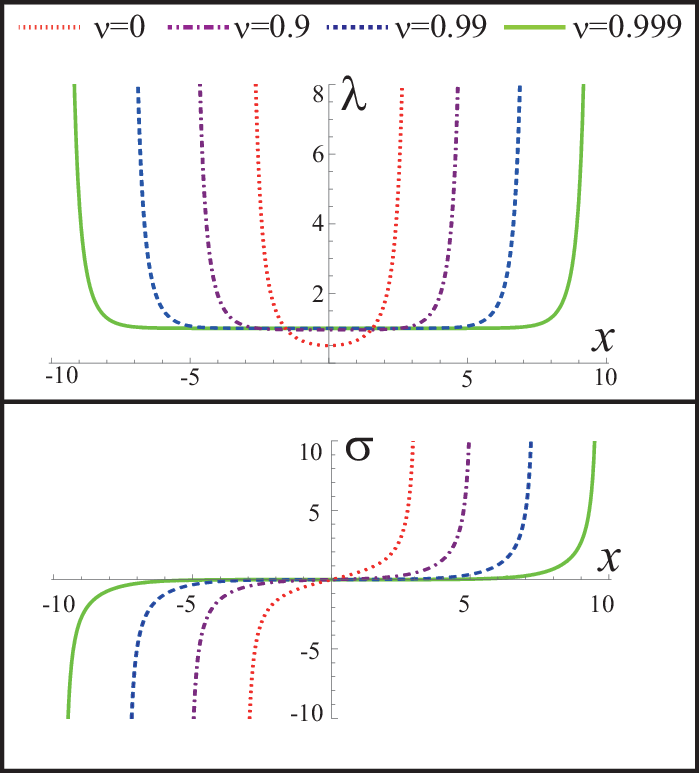}
\caption{The configuration of the mass gap function $\lambda$ and the Higgs field $\sigma$ for the confining phase. We set $\nu=0$ (dot line), $\nu=0.9$ (dash-dot line), $\nu=0.99$ (dash line), and $\nu=0.999$ (solid line). }
\label{Figlambdasigma}
\end{figure}

For the case of the infinite line, it has also been shown that the BCS solution corresponds to the confining phase, whereas the normal solution corresponds to the Higgs phase, which is inhibited by Coleman-Marmin-Wagner (CMW) theorem. 
In the case of infinite line, the gap equation (\ref{eq2}) has no physical solution due to the lack of an infrared cutoff for the normal solution \cite{Milekhin:2016fai} 
and it is consistent with the CMW theorem. 
Here, for the case of a  finite system, we shall refer to the phase corresponding to $\Delta_{\mathrm{BCS}}$ as the confining phase and we only consider the case of the confining phase. 
By using the mapping 
(\ref{lambdamap}) and (\ref{sigmamap}), 
we obtain exact self-consistent solutions 
\begin{eqnarray}
&&\lambda_{\mathrm{Conf}}=4\kappa^2
\frac{1-\mathrm{cn}(2\kappa x,\nu)\mathrm{dn}(2\kappa x,\nu)}{\mathrm{sn}^2(2\kappa x,\nu)},
\label{lambdaconf}
\\
&&\sigma_{\mathrm{Conf}}=A_{\mathrm{conf}} 
\frac{\mathrm{dn}(2\kappa x,\nu)-\mathrm{cn}(2\kappa x,\nu)}{\mathrm{sn}(2\kappa x,\nu)},
\label{sigmaconf}
\end{eqnarray}
for the confining phase. %and 

Since the confining solution corresponds to the usual confining solution in the infinite-length limit, where $\lambda(x)=\text{const.}$\ 
and $\sigma(x)=0$, one can show that the connection between the dynamical scale and the running coupling becomes 
\begin{align}
&\frac{N}{4\pi}\int_{-\Lambda_{\mathrm{UV}}}^{\Lambda_{\mathrm{UV}}} dk \frac{1}{\sqrt{k^2+\lambda}}-r_{\Lambda_{\mathrm{UV}}}=0\nonumber\\ 
&\Rightarrow\ \lambda=\Lambda_{\mathrm{UV}}^2\exp\left(-\frac{4\pi r_{\Lambda_{\mathrm{UV}}}}{N}\right) 
\end{align}
from Eq.\ (\ref{eq2}).

In Fig.\ \ref{Figlambdasigma}, we plot the mass gap function $\lambda$ and the Higgs field $\sigma$ 
for the confining phases. 
It can be seen from Fig.\ \ref{Figlambdasigma} that neither the mass gap function $\lambda$ nor the Higgs field $\sigma$ vanishes, whereas either $\lambda$ or $\sigma$ vanishes in the infinite system. 
Instead, both the mass gap function $\lambda$ and the Higgs field $\sigma$ diverge near the boundary for all the solutions, 
consistently with the numerical results in Ref.\ \cite{Bolognesi:2016zjp}. 
However, one can also see that the solutions 
behave as those for the infinite system except in the vicinity of the boundary. 
The divergence near the vicinity of the boundary stems from the infinite potential height of the model. 
Physically, this can be understood as follows. 
Since we have solved the problem in a self-consistent manner, the solution with the less singular behavior near the vicinity is chosen as an energetically favored solution. 
In fact, all the self-consistent potentials has $1/x$ and $1/(L-x)$ divergence near $x=0$ and $x=L$, respectively, which is much softer than the external potential $\infty\times \theta(-x)+\infty\times \theta(x-L)$ (Dirichlet boundaries). 
In the realistic system, the Dirichlet potential must have some energy cutoff and the divergences of $\lambda$ and $\sigma$ are also suppressed.  

It is fair to stress that the value of the parameter $\nu$ should be fixed by minimizing the effective action for a given choice of the length scale $L$ and of the parameter $r$ (in principle, a different choice of boundary conditions or the presence of external fields may also alter the value of $\nu$). In the present work we have treated $\nu$ as a free parameter and checked how the Casimir force changes as a function of $\nu$. However, an heuristic argument to fix the value of $\nu$ follow from the minimization of the value that the action, viewed as a function of $\nu$, takes at the solutions for specific choices of $L$ and $r$. A simple numerical calculation shows that for small $L$ the typically preferred value of $\nu$ is close to zero, while for large $L$ larger values of $\nu$ are preferred (and depend more dramatically on the value of $r$).

In the following we focus on the case of $L\rightarrow \infty$ and $\nu\rightarrow 1$. 
In this case, the solutions in Eqs.\ (\ref{lambdaconf}) and 
(\ref{sigmaconf}) reduce to the confining 
solutions previously obtained in an infinite system. 
We refer the value of $\sqrt{\lambda}$, which is constant in the infinite limit, as $m_{\infty}$. 
In the finite system, it is the size $L$ that works as infrared cutoff; 
thus the condition $L\rightarrow \infty$ assumed above should be considered as $L\gg 1/m_{\infty}$. 

%%%%%%%%%%%%%%%%%%%%%%%%%%%%
\section{Casimir force}
The vacuum energy is given by Eq.\ (\ref{energy}) and consists of two terms. 
The first corresponds to the usual Casimir contribution; 
the other is a contribution stemming from the condensate and that can be roughly considered as a classical background-field, though it is formed by the interacting bosons in a self-consistent manner. 

\begin{figure}
\center{\includegraphics[width=20pc]{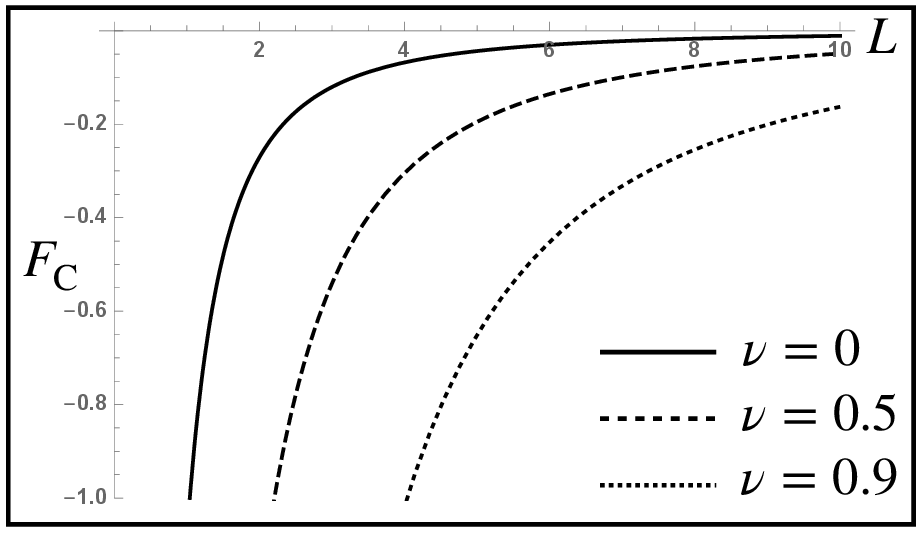}}
\caption{The Casimir force in the confinement phase for $\nu=0$, $0.5$, and $0.9$. The Casimir force is always attractive. We have set $r=1$. }
\label{FigFCConf}
\end{figure} 
\begin{figure}
\includegraphics[width=20pc]{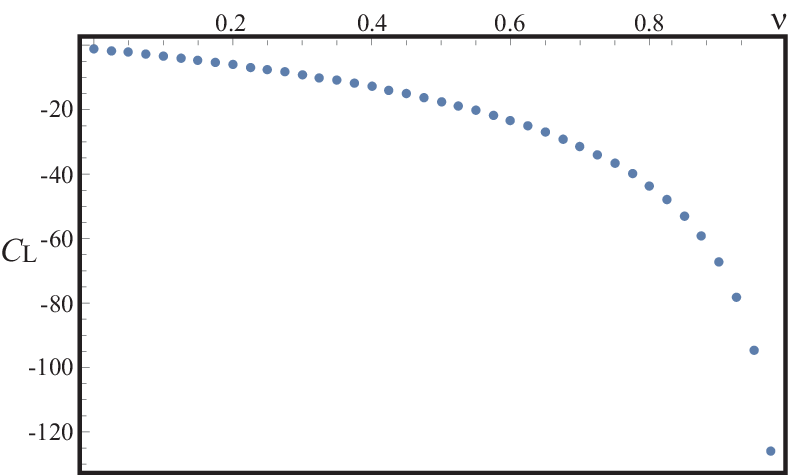}
\caption{The L\"usher coefficient $C_{\mathrm{L}}$ as a function of $\nu$. 
The coefficient exceed the value $-\pi/12$ corresponding to the noninteracting case. 
In the limit of $\nu\rightarrow 0$, $C_{\mathrm{L}}=-1.08333$. }
\label{FigLus}
\end{figure} 

For the potential (\ref{lambdaconf}), the energy spectrum is approximately given as $\omega_n=(\pi/L)\sqrt{(n+1)^2+m(n, \nu)^2}$. 
Numerical fitting indicates that the function $m(n, \nu)^2$ behaves as $m(n, \nu)=m(\nu)+\cdots$, 
consistently with the expected asymptotic behavior of the eigenvalues. 
In practise, we compute the eigenvalues numerically and fit, against the numerical data, the function $m(n, \nu) = m(\nu)+b(\nu)/ n+\cdots$. 
This ansatz allows us to control the diverging behavior of the Casimir energy and sub-leading terms of order O$(n^{-2})$) can be included systematically. 
Here we neglecting terms of order O$(n^{-2})$) and work at leading order. 
For regularization of the summation of the energies (first part of Eq.\ (\ref{energy})), we have employed the zeta function regularization where $\sqrt{n^2+m(n, \nu)^2}$ is analytically continued from $[(n+1)^2+m^2]^{s}$ with $s\to 1/2$. 
The other divergence from the condensation part (second part of Eq.\ (\ref{energy})) can be subtracted by changing the region to be $[\epsilon, L-\epsilon]$ and set $\epsilon\to 0$, where the $L$-independent diverging term must be subtracted.
The total energy is then computed numerically according to Eq.\ (\ref{energy}).
Illustrative results for the confining phase are shown in Fig.\ \ref{FigFCConf}. 
The Casimir force is always attractive. The confining phase is corresponding to the broken phase in the Gross-Neveu model and thus it may be expected that the sign change appears. 
This rather nontrivial situation is explained as follows. 
The bosonic spectrum which behaves $\sim \pi (n+1)/L$, instead of 
$\sim \pi (n+1/2)/L$ which appears for the fermionic system, 
results in the different behavior from the fermionic system.

The Casimir force is expected to have the L\"usher contribution 
which corresponds to the $C_{\mathrm{L}}/L^2$ term in Casimir force. 
In Fig.\ \ref{FigLus}, $\nu$ dependence of $C_{\mathrm{L}}$ is shown. 
It is known that the L\"usher term $-\pi/(12 L^2) $ appears in the Casimir force for the noninteracting case. 
In the presence of the Dirichlet boundary condition, the coefficient exceeds $-\pi/12$. 
Moreover, the strong enhancement is found by the many-body interaction.

\section{Summary}
In this work, we have considered the ${\mathbb C}P^{N-1}$ model on an interval as a prototype set-up of an interacting bosonic model.  
We have obtained  
exact self-consistent solutions for the case of Dirichlet boundary conditions imposed at the edges of the interval 
by using a recently discovered mapping between the 
Gross-Neveu model and the ${\mathbb C}P^{N-1}$ model \cite{Nitta:2017uog}. 
The solutions is specified by a continuous parameter $\nu\in [0,1]$. 
The solution reproduces the confinement phase in the infinite size limit ($\nu\rightarrow 1$ and $L=\infty$). 
We have found that both the mass gap function $\lambda$ and the Higgs field $\sigma$ diverge near the boundary for all the solutions, 
consistently with the numerical results in Ref.\ \cite{Bolognesi:2016zjp}. 
By using those solutions, we have calculated the vacuum energy and the Casimir force. 
The force is always attractive though the L\"usher term is strongly enhanced by the interaction. 

The results obtained here for the ${\mathbb C}P^{N-1}$ model (together with those of \cite{Schmidt}) and the result for the Gross-Neveu model \cite{Flachi:2017cdo} seem to indicate that the mechanism of inducing a change in the vacuum energy and in the force depends on the statistics. 

We conclude with few remarks and possible extensions. 
In the present work, we have focused on the Casimir effect for the confining phase of the ${\mathbb C}P^{N-1}$ model. 
Thus the existence of the Higgs phase in the finite system, which is inhibited in the infinite system, is still controversial. 
To substantiate the present results a more  
detailed analysis of the phase structure is necessary and we leave it 
for future work. 
However we point out the important feature which is expected if the Higgs phase exists. 
In the confining phase, the Higgs field $\sigma$ is an odd function and thus 
it cannot avoid to vanish at the origin, 
whereas it is always nonzero for the Higgs phase. 
Thus, the condition of $\sigma$ vanishing at some point may be used as 
a phase-identification criterion in numerical analyses. 

There is another issue on the inhomogeneous solutions. 
Recently there is a claim that the ground state is inhomogeneous even in the case of the infinite system \cite{Gorsky}.
In this paper, we have considered solutions in which the inhomogeneity appears only the vicinity of the boundaries. 
There should be a possibility that the solution with the inhomogeneity in the bulk could be energetically favorable. 
The ground state modulation in finite systems has also been discussed in the effective action approach in Ref.~\cite{Flachi:2019jus}. 
 Analytic self-consistent solutions for such a case may be also possible.

Investigating the nature of the transition characterizing the sign change in the force is an interesting problem that may unfold additional features on the mechanism controlling the sign flip.
Finding out whether a finite-size counterpart of the soliton lattice solution of Ref.\ \cite{Nitta:2017uog} exists is also an interesting question.
Looking at how external conditions like finite temperature or density alter the present results is certainly worth of considering and for such case alternative approaches (e.g., a l\'a Ginzburg-Landau) may be appropriate.
The supersymmetric ${\mathbb C}P^{N-1}$ model is another 
interesting direction due to additional cancellations that may occur in the vacuum energy due to supersymmetry. 
All such speculations are nontrivial. 

In closing we should mention that a related calculation has been carried out in Ref.\ \cite{Betti} where the focus is on the renormalized energy density rather than the Casimir force. The approach used there relies on numerics and their solutions differs from ours. This may be attributed to the differing boundary behaviour, and a closer comparison may shed some more light on the differences.

\section*{Acknowledgements}
The authors thank Kenichi Konishi and Keisuke Ohashi for informing us of their results. 
The authors gratefully acknowledge the Ministry of Education,
Culture, Sports, Science (MEXT)-Supported Program for the Strategic Research Foundation at Private Universities `Topological Science' (Grant No.~S1511006).
The work of M.~N.~is 
supported in part by the Japan Society for the Promotion of Science
(JSPS) Grant-in-Aid for Scientific Research (KAKENHI Grant
No.~16H03984 and No.~18H01217) 
and by a Grant-in-Aid for Scientific Research on Innovative Areas ``Topological Materials
Science'' (KAKENHI Grant No.~15H05855) from the MEXT of Japan. 
The work of R.~Y.~is  supported in part by the Japan Society for the Promotion of Science
(JSPS) Grant-in-Aid for Scientific Research (KAKENHI Grant No.~19K14616).

\end{document}